# Crystal orientation dependent spin pumping in $Bi_{0.1}Y_{2.9}Fe_5O_{12}$/Pt interface

Ganesh Gurjar[1,2,*], Vinay Sharma[3,4], Avirup De[5], Sunil Nair[5], S. Patnaik[1,*], Bijoy K. Kuanr[4,*]

[1]School of Physical Sciences, Jawaharlal Nehru University, New Delhi, INDIA 110067
[2]Department of Physics, Ramjas College, University of Delhi, New Delhi, INDIA 110007
[3]Department of Physics, Morgan State University, Baltimore, MD, USA 21251
[4]Special Centre for Nanosciences, Jawaharlal Nehru University, New Delhi, INDIA 110067
[5]Department of Physics, Indian Institute of Science Education and Research, Pune, INDIA 411007

## Abstract

Ferromagnetic resonance (FMR) based spin pumping is a versatile tool to quantify the spin mixing conductance and spin to charge conversion (S2CC) efficiency of ferromagnet/normal metal (FM/NM) heterostructure. The spin mixing conductance of FM/NM interface can also be tuned by the crystal orientation symmetry of epitaxial FM. In this work, we study the S2CC in epitaxial Bismuth substituted Yttrium Iron Garnet ($Bi_{0.1}Y_{2.9}Fe_5O_{12}$) thin films Bi-YIG (100 nm) interfaced with heavy metal platinum (Pt (8 nm)) deposited by pulsed laser deposition process on different crystal orientation $Gd_3Ga_5O_{12}$ (GGG) substrates i.e. [100] and [111]. The crystal structure and surface roughness characterized by X-Ray diffraction and atomic force microscopy measurements establish epitaxial Bi-YIG[100], Bi-YIG[111] orientations and atomically flat surfaces respectively. The S2CC quantification has been realized by two complimentary techniques, (i) FMR-based spin pumping and inverse spin Hall effect (ISHE) at GHz frequency and (ii) temperature dependent spin Seebeck measurements. FMR-ISHE results demonstrate that the [111] oriented Bi-YIG/Pt sample shows significantly higher values of spin mixing conductance (($2.31\pm0.23)\times10^{18}$ $m^{-2}$) and spin Hall angle ($0.01\pm0.001$) as compared to the [100] oriented Bi-YIG/Pt. A longitudinal spin Seebeck measurement reveals that the [111] oriented sample has higher spin Seebeck coefficient ($106.40\pm10$ nV $mm^{-1}$ $K^{-1}$). This anisotropic nature of spin mixing conductance and spin Seebeck coefficient in [111] and [100] orientation has been discussed using the magnetic environment elongation along the surface normal or parallel to the growth direction. Our results aid in understanding the role of crystal orientation symmetry in S2CC based spintronics devices.

**\*Corresponding authors:** ganeshgurjar4991@gmail.com, spatnaik@mail.jnu.ac.in, bijoykuanr@mail.jnu.ac.in

## 1. Introduction

Spin pumping is defined as the transfer of electron spins from a ferromagnet (FM) to a normal metal (NM) in an FM/NM junction under magnetic precession[1–3]. In general, a time varying magnetization pumps a pure spin current into a NM contact and this idea of spin pumping has been realized by numerous experimental techniques in which excitation sources are varied from GHz to THz time scales[4–6]. This is also observed with temperature gradient dependent spin Seebeck effect[7–9]. The efficiency of spin pumping relates to spin angular momentum transfer[10] and spin mixing conductance[11,12]. The nature of this efficiency is strongly affected by interface characteristics of NM in contact with either ferromagnetic metal (FM)[13] or ferrimagnetic insulator (FMI)[14]. A great deal of theoretical and experimental research has been focused on how to actuate, detect, and regulate the magnetization and spin currents in these systems[15]. One of the most important magnetic materials for spin pumping is Bismuth substituted Yttrium Iron Garnet (Bi-YIG, $Bi_xY_{3-x}Fe_5O_{12}$). This is primarily due to its long spin-wave propagation length and low Gilbert damping[16–22]. Several experimental studies on spin pumping have been reported using the Bi-YIG/Pt system[8,23–26] because of high spin orbit coupling (SOC) in Pt.

The recent theoretical and experimental works on this subject predict that the spin pumping efficiency strongly depends on the interface cut and orientation of NM and FM structures[27–31]. This anisotropy has been explained on the basis of local magnetic moment of magnetic ions and generation of crystal field symmetry due to broken rotational symmetry of magnetic atoms[27]. Motivated by these results, we investigate the role of crystal field symmetry in Bi-YIG/Pt multilayers. Here we discuss the tuning of spin pumping by controlling the interface cut dependent growth of 100 nm Bi-YIG film on $Gd_3Ga_5O_{12}$ (GGG) substrates. Our Bi-YIG/Pt heterostructures have been deposited using pulsed laser deposition (PLD) technique where GGG [100] and GGG [111] substrates are used to change the interface cut orientation. We utilize the broadband ferromagnetic resonance (FMR) technique to realize the spin pumping based on the enhancement of Gilbert damping and spin to charge conversion (S2CC) using inverse spin Hall effect (ISHE). The spin mixing conductance and spin Hall angle vary significantly with [111] and [100] orientations. To further corroborate our findings, we have used the complementary technique based on spin Seebeck effect where the variation of spin Seebeck voltage due to interface cut orientation has been observed.

## 2. Experimental Technique

The Bi-YIG thin films of thickness ≈100 nm were grown on [100] and [111] -oriented GGG substrates using PLD technique. Bare grown samples, GGG[100]/Bi-YIG and GGG[111]/Bi-YIG were labeled as Bi-YIG [100] and Bi-YIG [111] respectively. After that, a platinum thin film of ≈8 nm thickness with deposition rate of 0.8Å/sec was successfully deposited in-situ on bare Bi-YIG using PLD with optimized growth parameters. Platinum deposited samples, GGG[100]/Bi-YIG/Pt, and GGG[111]/Bi-YIG/Pt were labeled as Bi-YIG[100]/Pt and Bi-YIG[111]/Pt, respectively. The size of grown samples is ≈2 mm × 5 mm. Before deposition, GGG substrates were appropriately cleaned using acetone and isopropanol. All thin films were deposited at a base vacuum of $2\times10^{-7}$ mbar. We used a 248 nm KrF excimer laser with a 10 Hz pulse rate to ablate the material from the target. During deposition, Oxygen pressure, target to substrate distance, and substrate temperature were all maintained at 0.15 mbar, 5.0 cm, and 825 ºC, respectively. The growth rate of deposited Bi-YIG films was 6 nm/min. The as-grown films were in-situ annealed for 2 hours at 825 ºC in the presence of oxygen (0.15 mbar). After deposition of Bi-YIG film the target to substrate distance was changed to ≈3.0 cm and substrate temperature 100 ºC, for deposition of the Pt film, Thickness was measured using atomic force microscopy (AFM) (WITec GmbH, Germany). FMR-based spin pumping measurements have been carried out in a vector network analyzer spectrometer, where samples are placed in a flip-chip arrangement on a microstrip line with DC magnetic field applied perpendicular to the high-frequency magnetic field ($h_{RF}$) onto the film plane. The dc voltage generated due to ISHE was measured with the help of Keithley 2182 Nanovoltmeter. For spin Seebeck effect measurements, standard longitudinal spin Seebeck effect (LSSE) geometry was used with a temperature gradient formed between the Pt layer and the GGG substrate, and the voltage (perpendicular to an applied magnetic field and temperature gradient) was measured using a Keithley 2182 Nanovoltmeter.

## 3. Results and Discussion

### 3.1 Structural, surface morphology and static magnetization study

Bi-YIG crystallizes in cubic structure with the Ia-3d space group. Figure 1 (a)-(d) shows the XRD pattern of Bi-YIG and Bi-YIG/Pt bilayers on [100] and [111] cut GGG substrates. XRD patterns shown in Fig. 1 (a) and 1 (b) indicate the single-crystalline growth of Bi-YIG and Bi-YIG/Pt thin films. Figures 1 (c) and 1 (d) show the evidence for successful growth of Pt layer on top of bare Bi-YIG films of [100] and [111] orientations. The growth of Pt film is nano-crystalline in nature. The lattice constant, lattice mismatch (with respect to substrate), and lattice volume obtained from XRD data are listed in Table 1. The cubic structure lattice constant $a$ is calculated using the formula

$$a = \frac{\lambda\sqrt{h^2+k^2+l^2}}{2 \sin\sin\theta} \tag{1}$$

Where, $\lambda$ is the wavelength of Cu-K$\alpha$ radiation, $\theta$ is the diffraction angle, and [h, k, l] are the Miller indices of the corresponding XRD peaks.

The lattice mismatch ($\frac{\Delta a}{a}$) is calculated using the equation

$$\frac{\Delta a}{a} = \frac{(a_{film} - a_{substrate})}{a_{film}} \times 100 \tag{2}$$

Here, $a_{film}$ and $a_{substrate}$ indicate the lattice constants of the film and substrate, respectively. The calculated value of lattice constant are tabulated in Table 1 which are consistent with prior findings[16,21,32,33]. The lattice constant slightly increases in the case of [111] as compared to [100] because the distribution of $Bi^{3+}$ at the dodecahedral site is dependent on the substrate orientation[21,34,35] which leads to a comparatively larger lattice mismatch (or strain). Observed values of lattice mismatch are close to what has been reported earlier[36,37]. The lattice mismatch of the film plays an important role, a smaller lattice mismatch value can reduce the damping constant of the film[38,39].

Furthermore, the surface roughness plays an important role in magnetization dynamics due to the generation of two magnon scattering in films with higher roughness. Figure 1 (e)-(h) shows room temperature AFM images with root mean square (RMS) roughness. We have observed RMS roughness around 0.35 nm or less for all grown Bi-YIG films which is comparable to previously reported YIG films[18,40]. We have observed that there RMS roughness remains unchanged for all grown films. Moreover, roughness would be more affected by changes in growth factors than by substrate orientation[18,21].

VSM magnetization measurements were performed at 300 K with an applied magnetic field parallel to the film plane (in-plane). The GGG substrate's paramagnetic contribution was properly subtracted. Magnetization plots of Bi-YIG thin films for [100] and [111] orientations are shown in Figure 1(i). The schematic of an applied magnetic field direction parallel to the film plane is shown in the inset of Fig. 1(i). Measured saturation magnetization ($\mu_0 M_S$) values are 165.60 ± 20.10 mT Gauss for Bi-YIG [100] and 196.20 ± 19.10 mT for Bi-YIG [111]. Saturation magnetization error bars are related to sample volume uncertainty. The observed value of $\mu_0 M_S$ of as grown [100] and [111]-oriented Bi-YIG films are consistent with previous reports[33,41–43]. Bismuth doping generally affects the magnetic and mechanical properties of YIG, However, the dependence of saturation magnetization on interface cut is minimal because the magnetization of YIG is induced via a super-exchange interaction at the *d* and *a* site (in Wycoff notation) between non-equivalent $Fe^{3+}$ ions. Along [111], there is more contribution of $Bi^{3+}$ ions compared to [100] orientation[21], and Bismuth located at the dodecahedral site distort the tetrahedral and octahedral $Fe^{3+}$ ions[22]. Figure 1 (j) shows the crystal orientation dependent atomic model of YIG with Yttrium (Y) (red), Fe 'd site' (green) and Fe 'a site' (blue) atomic position. The interface cut dependent magnetic environment is highly anisotropic in [111] and [100] orientation which is clearly depicted in Fig. 1(j). Pt interface with these different magnetic environments may change its spin conversion efficiency[44] which will be discussed in next sections.

**Table 1:** Lattice constant and roughness obtained from XRD and AFM.

| S. No. | Sample | Lattice constant (Å) | Lattice Mismatch (%) | Lattice volume (Å$^3$) | Roughness (nm) |
|---|---|---|---|---|---|
| 1. | Bi-YIG [100] | 12.461 | 0.53 | 1935.15 | 0.36 |
| 2. | Bi-YIG [111] | 12.472 | 0.64 | 1939.80 | 0.33 |
| 3. | Bi-YIG/Pt [100] | 12.402 | 0.39 | 1907.81 | 0.29 |
| 4. | Bi-YIG/Pt [111] | 12.455 | 0.80 | 1932.31 | 0.34 |

### 3.2 Ferromagnetic resonance (FMR) study

FMR measurements are performed at room temperature. Figure 2 (a) and (b) show the FMR absorption spectra of [100] and [111] -oriented thin films, respectively. When Pt is deposited on top of Bi-YIG, the observed FMR absorption spectra shift right in case of [100] and left in case of [111] orientation with respect to bare Bi-YIG film. This shows that Pt deposition moderately affects the effective saturation magnetization in Bi-YIG[111] due to its [111] texture. The FMR data at 10 GHz with its Lorentzian fit is shown in the inset of Fig. 2 (a) and 2 (b), and the schematic depicts the direction of the applied DC magnetic field in the film plane. Inset in Fig. 2 (a) and (b) shows the enhancement of FMR linewidth in Bi-YIG/Pt samples as compared to bare Bi-YIG samples. The corresponding FMR linewidth enhancement is ~0.24 mT in [100] and ~0.50 mT in [111] oriented films. This enhancement in linewidth results from spin pumping. The coupling that transfers angular momentum from Bi-YIG to the metal contributes to the damping due to precession of Bi-YIG magnetization that leads to increasing linewidth. To calculate the increment in Gilbert damping due to Pt deposition, we have measured the FMR data at different resonance frequencies ($f$ = 2 GHz-12 GHz) with dc magnetic field applied parallel to film plane. The Lorentzian fitting was used to fit the FMR data and calculate the FMR linewidth (ΔH) and resonance magnetic field ($H_{rf}$) or different frequencies. From the fitting of Kittel's in-plane equation provided by Eq. 3[45], the gyromagnetic ratio (γ) and effective magnetization field ($\mu_0 M_{eff}$) were calculated.

$$f = \frac{\gamma}{2\pi} \mu_0 \sqrt{(H_r)(H_r + M_{eff})} \qquad (3),$$

Where, $\mu_0 M_{eff} = \mu_0 (M_s - H_{ani})$ is the effective field with anisotropy field $H_{ani} = \frac{2K_1}{M_s}$. Similarly, Gilbert damping parameter (α) and inhomogeneous broadening (ΔH$_0$) linewidth were calculated from the fitting of the Landau–Lifshitz–Gilbert equation (LLG) provided by equation 4[45],

$$\Delta H(f) = \Delta H_0 + \frac{4\pi\alpha}{\gamma} f \qquad (4)$$

Figures 2 (c) and 2 (d) show Kittel and LLG fitted graphs for the [111] orientation, respectively, with an inset showing Kittel and LLG fitted graphs for the [100] orientation.

Table 2 lists the calculated parameters obtained from the FMR study. The measured Gilbert damping (α) is consistent with thin films used in spin-wave propagation studies[21,24,43]. For both orientations [100] and [111], the value of α increases when Pt is deposited on top of Bi-YIG films (Table 2). The value of α increases significantly in (111) orientation. However, in the case of pure YIG it is reported that Gilbert damping in [100] is more as compared to [111][28]. In our case higher α obtained in [111] may be ascribed qualitatively to the presence of $Bi^{3+}$ ions, which additional induce spin-orbit coupling (SOC)[16,46,47] that causes local distortion of $Fe^{3+}$ so as to affect the magnetic properties as compared with pure YIG[22]. The higher lattice mismatch (strain) in (111) could also be the reason for increased electron scattering leading to higher damping[48]. Figure 1 (j) elucidates how more Fe atoms (i.e., more magnetic environment) and also more yttrium sites are available for the distribution of $Bi^{3+}$ ions along [111] orientation. This is the cause of higher lattice mismatch in Bi-YIG [111]. These findings account for the higher Gilbert damping and $\mu_0 M_{eff}$ values in the [111] orientation. In conclusion, Bi-YIG/Pt with an orientation of [100] has the lowest damping factor. We note that these are the essential desirable factors used in spintronic devices that need longer propagation lengths for spin waves.

**Table 2:** Damping and linewidth parameters obtained from FMR

| S. No. | Sample | α (×$10^{-4}$) | Δ$H_0$ (Oe) | $\mu_0 M_{eff}$ (Oe) |
|---|---|---|---|---|
| 1. | Bi-YIG [100] | (2.97±0.11) | 25.99±0.21 | 2124.89±8.80 |
| 2. | Bi-YIG [111] | (3.62±0.17) | 26.69±0.33 | 2241.70±12.77 |
| 3. | Bi-YIG/Pt [100] | (3.74±0.14) | 27.33±0.29 | 2030.12±40.19 |
| 4. | Bi-YIG/Pt [111] | (6.11±0.47) | 25.19±0.92 | 2582.26±75.13 |

## 3.3 Inverse spin Hall effect (ISHE) study using FMR based spin pumping and spin Seebeck effect

The induced crystal field anisotropy due to different interfaces strongly affects the tuning of spin mixing conductance because the magnetic $Fe^{3+}$ ions can result in different spin currents according to its placement in [111] and [100] orientation. It is argued that the crystal field is elongated along the surface normal in [111] orientation and parallel to the surface in [100] orientation[27]. This anisotropic nature of the crystal field gives the highest spin mixing conductance in [111] direction and considerably lower in [100] orientation. At Bi-YIG/Pt interfaces where the spin pumping or the transfer of spin angular momentum occurs, the spin mixing conductance ($g_{\uparrow\downarrow}$) plays a significant role in transfer of spin angular momentum. Spin pumping-induced damping is measured from the enhancement in the Gilbert damping constant (e.g. due to Pt layer on top of Bi-YIG film ($\alpha_{Bi-YIG/Pt} - \alpha_{Bi-YIG}$)). This is required for calculating the spin mixing conductance at the Bi-YIG/Pt interface. Table 2 shows the Gilbert damping values of bare Bi-YIG and Pt deposited Bi-YIG samples. The value of $g_{\uparrow\downarrow}$ as (0.73±0.07) ×$10^{18}$ m$^{-2}$ for [100] and (2.31±0.23) ×$10^{18}$ m$^{-2}$ for [111] oriented samples are calculated using equation 5[14]. The observed spin mixing conductance values are consistent with previous reports[14,24,49,50] and shows that it depends on crystal orientation. A larger spin mixing conductance in [111] oriented samples occur that indicates that [111] grown samples have a greater spin-injection efficiency than [100] oriented samples. This is a consequence of [111] oriented sample having a greater Gilbert damping constant enhancement owing to the Pt layer.

$$g_{\uparrow\downarrow} = \frac{\mu_0 M_s t_F}{g\mu_B}\left(\alpha_{Bi-YIG/Pt} - \alpha_{Bi-YIG}\right) \qquad (5)$$

Where $\mu_0 M_s$ and $t_F$ are the magnetization and thickness of the Bi-YIG thin film, $\mu_B$ is the Bohr magneton, $g = \frac{\gamma\hbar}{\mu_B}$ is the g-factor of Bi-YIG[22] and $\alpha_{Bi-YIG/Pt} - \alpha_{Bi-YIG}$ is the enhancement in the Gilbert damping constant due to Pt layer on top of Bi-YIG film.

ISHE measurements are carried out at room temperature using FMR technique. In this scheme, a dc magnetic field (H) is applied in the film plane and perpendicular to the sample length. ISHE voltage (V$_{ISHE}$) is measured along the length of the Pt layer. A resonant rf field (h$_{rf}$), is also superimposed to cause the uniform precession of the Bi-YIG magnetization. When excited precession occurs at the interface, angular momentum is transferred to the conduction electrons in Pt, resulting in a pure spin current (J$_S$) in Pt. As a result of ISHE, the spin current travels across the Pt perpendicular to the plane and is transformed into charge current (J$_C$). As a consequence of this, a voltage difference develops between the two ends of the Pt layer that is measured using a Keithley nanovoltmeter. Figure 3 (a) shows the measured dc voltage (V$_{dc}$) at frequency 10 GHz, and figure 3 (b) shows its Lorentzian fitting[45] with symmetric and anti-symmetric contributions. Figure 3 (c) shows the symmetric contribution of the voltage (V$_{ISHE}$) extracted from the measured dc voltage (V$_{dc}$). The inset shows the schematic setup for FMR spin pumping for measurement of ISHE voltage.

The ISHE voltages[14] given by equation 6 depends on several material parameters

$$V_{ISHE} = \frac{-e\theta_{SH}}{\sigma_N t_N + \sigma_F t_F} \lambda_{SD} \tanh\left(\frac{t_N}{2\lambda_{SD}}\right) g_{\uparrow\downarrow} fLP\left(\frac{\gamma h_{rf}}{2\alpha\omega}\right)^2 \qquad (6)$$

Where e is the electron charge, $\theta_{SH}$ is the spin Hall angle, $\sigma_N(\sigma_F)$ and $t_N(t_F)$ denotes the conductivity and thickness of the NM (FM) thin-film respectively, $\lambda_{SD} = 7.3\ nm$ is the spin diffusion length in Pt[16], $g_{\uparrow\downarrow}$ denotes interfacial spin mixing conductance, $\omega = 2\pi f$ is the resonance frequency, L is the length of the sample, and $h_{rf} = 0.17\ Oe$ in our FMR cavity at power = +15 dBm. The ellipticity of the magnetization precession gives rise to the factor $P$[13,14].

$$P = \frac{2\omega[\gamma\mu_0 M_s + \sqrt{(\gamma\mu_0 M_s)^2 + 4\omega^2}]}{(\gamma\mu_0 M_s)^2 + 4\omega^2} = 1.21 \tag{7}$$

Values of $g_{\uparrow\downarrow}$ and $\theta_{SH}$ are calculated using equations (5)-(7). Calculated values of $\theta_{SH}$ for [100] orientation is $(0.73\pm0.07)\times10^{-2}$ and for [111] orientation is $(1.01\pm0.10)\times10^{-2}$. Figure 3 (d) shows the comparative results of $g_{\uparrow\downarrow}$ and $\theta_{SH}$ for [100] and [111] oriented samples. Consequently, the spin current density $J_s$ can be calculated using[14],

$$J_s = \frac{\sigma_N t_N + \sigma_F t_F}{\theta_{SH}\lambda_{SD}\tanh\left(\frac{t_N}{2\lambda_{SD}}\right)} \frac{V_{ISHE}}{L} \tag{8}$$

$J_s$ is calculated to be $0.59\times10^7$ A-m$^{-2}$ for [100] and $0.70\times10^7$ A-m$^{-2}$ for [111] orientation. $g_{\uparrow\downarrow}$, $\theta_{SH}$ and $J_s$ was significantly larger in the case of [111] orientation, indicating the enhanced spin orbit contribution and spin transparency in [111] oriented Bi-YIG/Pt bilayer structure, resulting in maximum $V_{ISHE}$ signal ~76.30 μV. For the [100] oriented Bi-YIG/Pt film it was found to be ~46.31 μV. Our obtained transport parameters are in agreement with the reported literature values[14,51–58] in which order of spin mixing conductance values varies from $\approx 10^{16}$ m$^{-2}$ to $10^{18}$ m$^{-2}$. Similarly order of spin Hall angle[57] and spin current density are also in agreement with published data[27,29–31]. It shows the effect of crystal orientation on spin mixing conductance. Jia et al. [31] observed no crystal orientation dependent $g_{\uparrow\downarrow}$ in the pure YIG/Ag system. Cahaya et al.[27], reported that asymmetry of spin mixing conductance depends not only on the density of exposed moments and but also on local point symmetry. The pumped spin-current anisotropy is dependent on the quadrupole moment, which is dependent on the orbital occupancy of interface magnetic atoms. For half-filled shells ions ($Fe^{3+}$) quadrupole moment is zero. It is necessary to break symmetry, such as via strain or at an interface, to obtain a finite quadrupole moment. From the theoretical explanation given by[27], we conclude that our crystal orientation dependent findings may be due to rotational symmetry breaking in Bi-YIG thin films owing to substrate lattice mismatch. As discussed already, a larger lattice mismatch in [111] is observed compared to [100]. Furthermore, we have used Bi-doped YIG system, where $Bi^{3+}$ affects the magnetic properties of YIG[22] and also distribution of $Bi^{3+}$ ions among dodecahedral sites depends on substrate orientation[21,59–61]. Figure 1(j) shows the interface cut dependent magnetic environment in [100] and [111] orientation. The interaction of Pt with these varied magnetic environments may affect its spin conversion efficiency. Hence, the enhancement in spin pumping parameters depends on the orientation of crystal growth[27,29–31], which supports our experimental findings of ISHE results.

**3.4 Spin Seebeck effect**

The integration of magnetic insulators with Pt promises to usher in several applications of spin current technology to thermoelectric devices. In this regard, we discuss spin Seebeck study on the same films that were studied for inverse Spin Hall Effect previously. We note that the doping of Bi softens the crystal and increases the growth induced anisotropy[22]. The effective SOC for $Fe^{3+}$ ions is increased by mixing 6p orbitals of Bi ions with O-2p orbitals. Energy transmission between spin and phonon systems is also enhanced by SOC. Thus during spin thermoelectric generation, maximum energy transfer mechanism for the heat simulation of magnetization precession will result in larger spin currents in Bi-YIG films[23]. The spin Seebeck effect measurements were performed in longitudinal spin Seebeck effect (LSSE) geometry. Figure 4 (a) shows the schematic setup for LSSE measurements. Figure 4 (b), 4 (c), and 4 (d) show the comparative results of longitudinal spin Seebeck voltage ($V_{LSSE}$) as a function of temperature (20 K to 300 K), applied magnetic field (up to ±0.2 Tesla) and temperature gradient (upto 11 K) respectively. At room temperature, the observed spin Seebeck coefficient is 106.40±10 nV mm$^{-1}$ K$^{-1}$ for [111] and 89.41±0.3 nV mm$^{-1}$ K$^{-1}$ for [100] oriented sample. Because these voltages are obtained across various effective resistances, we represent these voltages in their normalized form, i.e. $V_{LSSE} = V_{dc}/R\Delta TL$, where $V_{dc}$ is the observed transverse voltage, R and L are the corresponding resistance and distance between contact probes, $\Delta T$ is the temperature gradient across the sample. The Spin Seebeck effect study indicates the significantly higher values of spin Seebeck coefficient observed in the case of the [111]. These findings from spin Seebeck measurements also provide credence to the higher ISHE voltage generated in the [111] orientation that was seen in FMR spin pumping studies.

According to these experimentally observed ISHE results, the spin mixing conductance of Bi-YIG/Pt bilayers depends on crystal cut and orientation[27]. We find that the spin pumping efficiency of the [111] oriented sample is substantially greater than that of the [100] orientated sample.

## 4. Conclusion

We have successfully grown the Bi-YIG(100 nm)/Pt(8 nm) bilayer using pulsed laser deposition technique on top of GGG substrates having [100] and [111] orientations. AFM and XRD characterizations revealed that the deposited thin films are phase pure and have smooth surfaces. FMR based spin pumping results confirm that there are significantly higher values of spin mixing conductance (~ three times) and spin Hall angle in the case of the [111] orientated sample compared to [100] oriented sample. This indicates that significantly higher spin current can transfer to the Pt from Bi-YIG [111]. Consequently, the [111] oriented sample has a higher spin pumping efficiency than the [100] oriented sample. From a longitudinal spin Seebeck measurement we note that the [111] orientated sample had a higher (~20 %) spin Seebeck coefficient. In conclusion, the results of inverse spin Hall effect and spin Seebeck experiments show that the fundamental parameters of spin pumping can be tuned effectively by substrate's orientation.

**Acknowledgements**

This work is supported by the MHRD-IMPRINT grant, DST (SERB, AMT, and PURSE-II) grant of Govt. of India. Ganesh Gurjar acknowledges CSIR, New Delhi for financial support. We acknowledge AIRF, JNU for access of PPMS facility.

**Conflict of interest statement**

The authors declare that they have no competing financial interests or personal relationships that could have appeared to influence the work reported in this paper.

**Data Availability statement**

The data that support the findings of this study are available from the corresponding author upon reasonable request.

**Figure Captions**

**Figure 1:** XRD patterns of GGG/Bi-YIG/Pt structure, Bi-YIG XRD pattern shown (a) with [100] and (b) [111] orientations, (c) and (d) shows the XRD pattern of Pt film with respect to bare Bi-YIG films on [100] and [111] orientations, (e)-(h) AFM images of Bi-YIG and Bi-YIG/Pt films in [100] and [111] orientations, (i) VSM plot of Bi-YIG film in [100] and [111] orientations (one inset shows the schematic of applied magnetic field direction in the film plane, second inset shows the schematic of sample grown structure), (j) shows the crystal orientation dependent atomic model of YIG with Yttrium (Y) (red), Fe 'd site' (green) and Fe 'a site' (blue) atomic position.

**Figure 2:** FMR absorption spectra of Bi-YIG and Bi-YIG/Pt films with (a) [100] and (b) [111] orientation. Inset shows FMR data at 10 GHz with its Lorentzian fit and enhancement in FMR linewidth in Pt deposited sample. The schematic shows the direction of the applied DC magnetic field in the film plane. (c) and (d) show frequency-dependent FMR field ($H_r$) data fitted with Kittel equation and frequency-dependent FMR linewidth ($\Delta H$) data fitted with LLG equation [111] orientation, respectively, with an inset showing Kittel and LLG fitted graphs for the [100] orientation.

**Figure 3:** (a) show the measured dc voltage ($V_{dc}$) at frequency 10 GHz, (b) show its Lorentzian fitting with symmetric and anti-symmetric contributions, (c) symmetric contribution of the voltage ($V_{ISHE}$) extracted from the measured dc voltage ($V_{dc}$) (schematic setup of ISHE measurements), (d) shows the comparative results of spin mixing conductance ($g_{\uparrow\downarrow}$) and spin Hall angle ($\theta_{SH}$) for [100] and [111] oriented samples.

**Figure 4:** (a) Shows the schematic setup of spin Seebeck measurements in LSSE geometry, comparative results of longitudinal spin Seebeck voltage ($V_{LSSE}$) as a function of (a) ambient temperature, (b) applied magnetic field, (c) temperature gradient, for [100] and [111] oriented samples.

**Figure 1:**

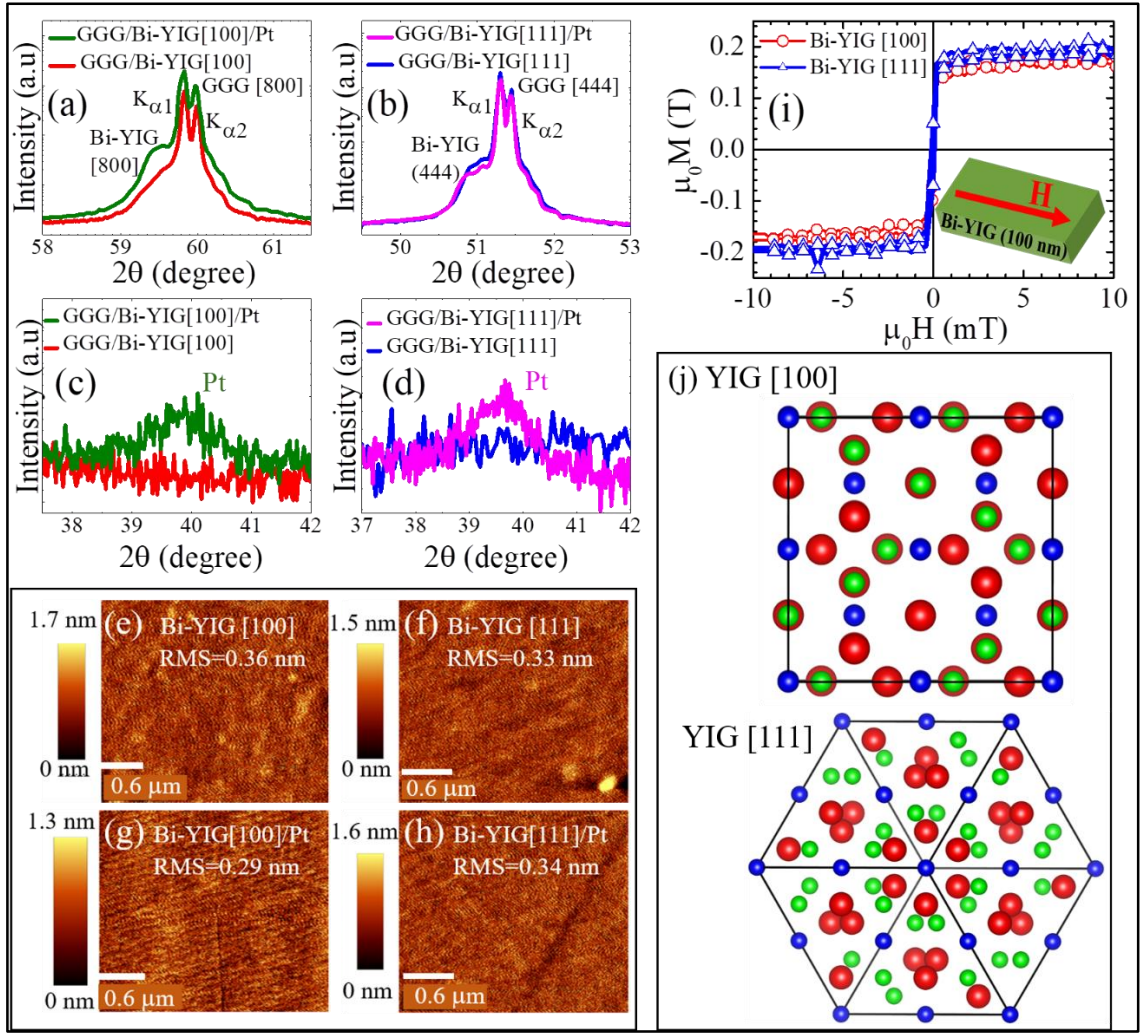

**Figure 2:**

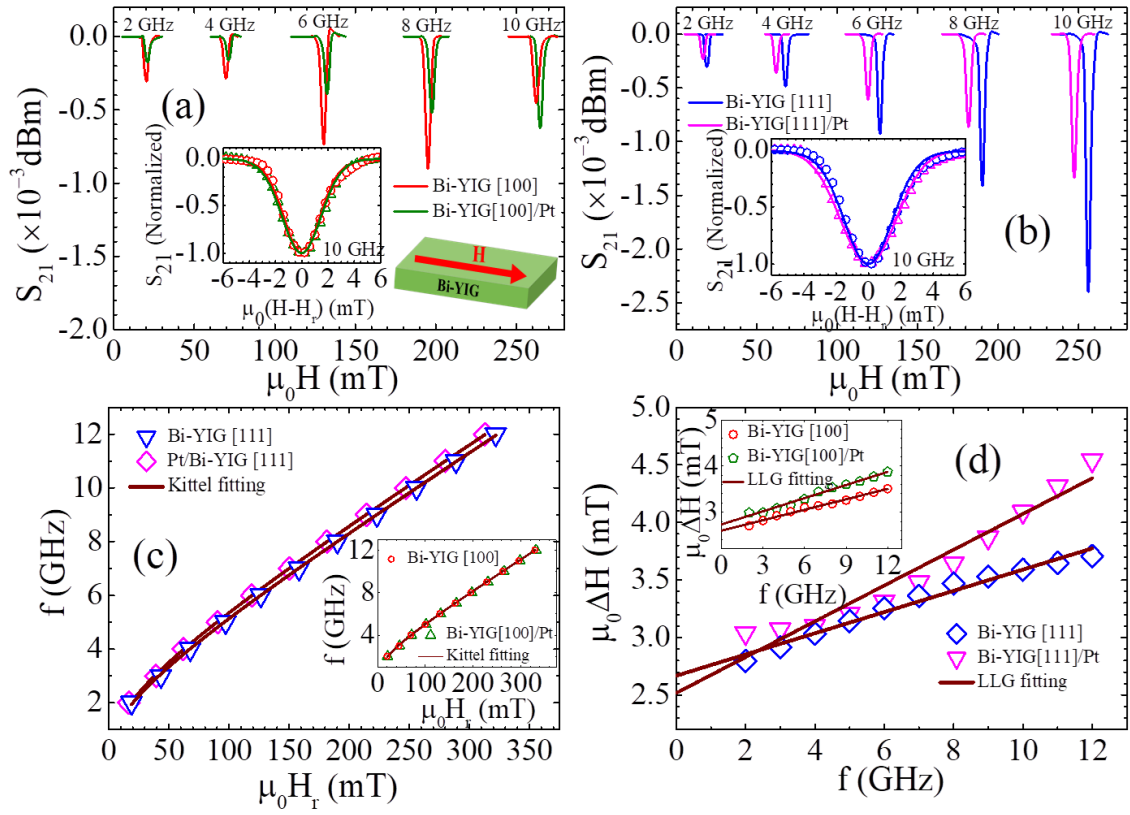

**Figure 3:**

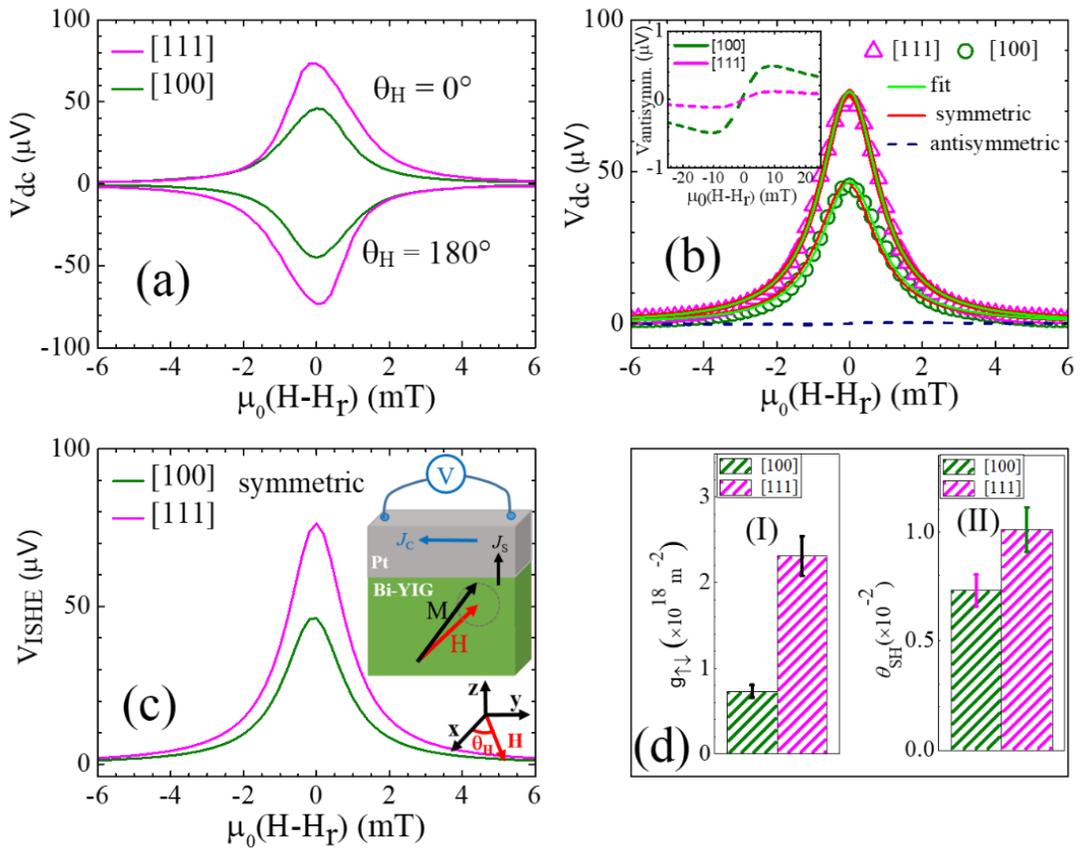

**Figure 4:**

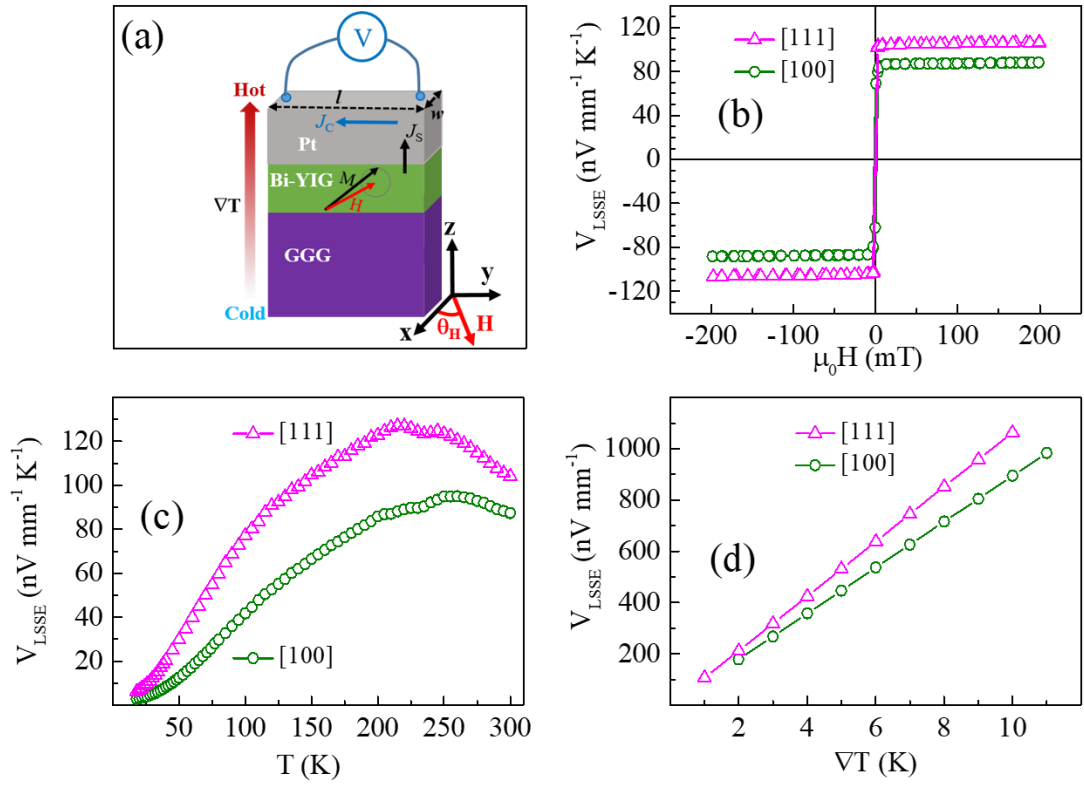